\documentstyle[12pt,aasms4]{article}
\def\etal{{\it et~al.\ }}
\def\sles{\lower2pt\hbox{$\buildrel {\scriptstyle <}
	\over {\scriptstyle\sim}$}}
\def\sgreat{\lower2pt\hbox{$\buildrel {\scriptstyle >}
	\over {\scriptstyle\sim}$}}

\received{................}
\accepted{................}
\begin{document}

\title{The Nature of Compact Galaxies in the Hubble Deep Field --- II: \\
Spectroscopic Properties and Implications for the Evolution of the
Star Formation Rate Density of the Universe. \altaffilmark{1,2,3} }

\author{Rafael Guzm\'an, Jes\'us Gallego, David C. Koo, Andrew C. Phillips, 
 James D. Lowenthal\altaffilmark{4}, 
S. M. Faber, Garth D. Illingworth \& Nicole P. Vogt}
 
\affil{University of California Observatories / Lick Observatory, \\
Board of Studies in Astronomy and Astrophysics, \\
University of California, Santa Cruz, CA 95064}

\dates

\bigskip 
\bigskip
\bigskip

\centerline{To be submitted to {\it Astrophysical Journal}}

\altaffiltext{1}{Based on observations obtained at the W. M. Keck Observatory,
which is operated jointly by the University of California and the
California Institute of Technology.}

\altaffiltext{2}{Based in part on observations with the NASA/ESA {\sl
Hubble Space Telescope}, obtained at the Space Telescope Science
Institute, which is operated by AURA under NASA contract NAS 5-26555.}

\altaffiltext{3}{Lick Observatory Bulletin No.000}

\altaffiltext{4}{Hubble Fellow}

\newpage 

We present a spectroscopic study of 51 compact field galaxies with
redshifts $z < 1.4$ and apparent magnitudes I$_{814} <$ 23.74 in the
flanking fields of the Hubble Deep Field. These galaxies are compact
in the sense that they have small apparent half-light radii ($r_{1/2}
\le 0.5$ arcsec) and high surface brightnesses ($\mu_{I814} \le 22.2$
mag arcsec$^{-2}$).  The spectra, taken at the Keck telescope, show
emission lines in 88\% of our sample, and only absorption lines in the
remaining 12\%.  Emission-line profiles are roughly Gaussian with
velocity widths that range from the measurement limit of $\sigma \sim$
35 km s$^{-1}$ to 150 km s$^{-1}$. Rest-frame [OII]$\lambda$3727
equivalent widths range from 5\AA\ to 94\AA , yielding star formation
rates (SFR) of $\sim$0.1 to 14 M$_\odot$ yr$^{-1}$.  The analysis of
various line diagnostic diagrams reveals that $\sim$60\% of compact
emission-line galaxies have velocity widths, excitations, H$\beta$
luminosities, SFRs, and mass-to-light ratios characteristic of young
star-forming HII galaxies. The remaining 40\% form a more
heterogeneous class of evolved starbursts, similar to local starburst
disk galaxies. We find that, although the compact galaxies at $z>0.7$
have similar SFRs per unit mass to those at $z<0.7$, they are on
average $\sim$10 times {\sl more massive}.  Our sample implies a lower
limit for the global comoving SFR density of $\sim$0.004 M$_\odot$
yr$^{-1}$ Mpc$^{-3}$ at $z =$ 0.55, and $\sim$0.008 M$_\odot$
yr$^{-1}$ Mpc$^{-3}$ at $z =$ 0.85 (assuming Salpeter IMF, H$_0 =$ 50
km s$^{-1}$ Mpc$^{-1}$, and q$_0 =$ 0.5).  These values, when compared
to estimates for a sample of local compact galaxies selected in a
similar fashion, support a history of the universe in which the SFR
density declines by a factor $\sim$10 from $z =1$ to today. From the
comparison with the SFR densities derived for magnitude-limited
samples of field galaxies, we conclude that compact emission-line
galaxies, though only $\sim$20\% of the general field population, may
contribute as much as $\sim$45\% to the global SFR of the universe at
$0.4 < z < 1$.

\vspace{-0.5cm}

\clearpage
\section{Introduction}

Faint compact galaxies are relevant to observational cosmology because
they serve to constrain several proposed explanations of the abundance
of faint blue field galaxies (see reviews by Koo and Kron 1992; Lilly
1993; Ellis 1996). These include models with large populations of
low-luminosity dwarfs at low redshifts (Phillipps \& Driver 1995);
bursting dwarfs at $z <~ 1$ that have faded or disappeared by today
(Cowie, Songaila \& Hu 1991; Babul \& Ferguson 1996); low-luminosity
AGN's (Tresse \etal 1996); or smaller pre-merger components
(Guiderdoni \& Rocca-Volmerange 1990; Broadhurst \etal 1992; Kauffmann
\etal 1993; Cole \etal 1994). Faint compact galaxies are also likely
to include compact narrow emission-line galaxies (CNELGs), which are
starbursts at moderate redshifts that have been proposed to be
progenitors of today's spheroidal galaxies like NGC 205 (Koo \etal
1994, 1995; Guzm\'an \etal 1996). Given their likely starburst nature,
faint compact galaxies may be major contributors to the global star
formation rate (SFR) density already found to increase with lookback
time to at least redshift $z \sim 1$ (Cowie et al. 1995, Lilly \etal
1996).

	 Most compact galaxies at moderate redshifts yield little
morphological information, even in HST images. Their integrated
spectra are thus particularly valuable in providing information on
their galaxy type, kinematics, and stellar content, as well as the
physical conditions of the ionized gas and star formation activity. In
this pair of papers, we study the properties of a sample of 51 faint
compact galaxies in the flanking fields of the Hubble Deep Field (HDF;
Williams \etal 1996).  The global properties of our sample are
described by Phillips \etal (1997; hereafter Paper I), who find that
the majority of faint compact galaxies tends to have colors,
luminosities, half-light radii, surface brightnesses, and
mass-to-light ratios consistent with those of local vigorously
star-forming galaxies.  In this paper, we focus our analysis on their
spectroscopic properties.  The wide spectral range (4000-9000\AA ) and
good resolution ($\sim$4\AA\ FWHM) of our survey offer significant
advantages for a spectral analysis of compact field galaxies at
intermediate redshifts. Most spectroscopic surveys of faint galaxies
do not cover the entire optical wavelength range, hence emission-line
studies have been restricted mainly to the [OII]$\lambda$3727 doublet
(e.g., Broadhurst \etal 1988; Colless \etal 1990; Glazebrook \etal
1995).  At $z<0.7$, we can reliably measure not only
[OII]$\lambda$3727 but also [OIII]$\lambda$4959, [OIII]$\lambda$5007,
and H$\beta$. Flux ratios among these lines provide powerful
diagnostics to discriminate among different classes of emission-line
galaxies (Baldwin, Phillips \& Terlevich 1981; Veilleux \& Osterbrock
1987). For example, a recent study of line ratios for field galaxies
at $z<0.3$ suggests that at least 8\% are active galaxies such as
Seyfert 2 or LINERS (Tresse \etal 1996). A second major distinguishing
feature of the present study is the inclusion of velocity width
measurements from emission lines.  Internal velocities, as inferred
from the motions of the ionized gas, have proven to be very useful in
understanding the nature of distant field galaxies and assessing their
evolutionary state (Koo \etal 1995; Forbes \etal 1996; Guzm\'an \etal
1996; Vogt \etal 1996; Rix \etal 1997). Together with other
spectroscopic parameters such as the excitation of the ionized gas,
H$\beta$ luminosity, and current SFR, the new velocity widths improve
discrimination among different classes of emission-line galaxies and
provide a more reliable comparison to potential local counterparts.

	This paper is organized as follows. In Section 2 we describe
briefly our sample selection and spectroscopic observations. In
Section 3 we describe the emission-line measurements. In Section 4 we
study the nature of the compact emission-line galaxies. In Section 5
we derive the comoving SFR density at $0.4 \le z \le 1.0$ for our
sample and compare our results to previous observations and model
predictions. The main results of this paper are summarized in Section
6. Throughout this paper we adopt H$_0 =$ 50 km s$^{-1}$ Mpc$^{-1}$
and q$_0 =$ 0.05, unless otherwise stated. Given these parameters,
$L^\star$ ($M_B \sim -21$) corresponds to $I_{814} \sim 21$ and
$1''$ spans 9 kpc at a redshift of $z=0.7$. This project is part of
the DEEP program (Koo 1995).

\section{Description of the sample and observations} The galaxy sample
was selected from I$_{814}$ HST images of the flanking fields around
the HDF (Williams \etal 1996).  A full description of the sample
selection, spectroscopic observations, and measurements of photometric
parameters is given in Paper I.  Briefly, these galaxies were chosen
to have I$_{814} \le 23.74$, half-light radius $r_{1/2} \le 0.5$
arcsec, and average surface brightness within the half-light radius
$\mu_{I814} \le 22.2$ mag arcsec$^{-2}$.  Hereafter we refer to this
sample as {\sl compact galaxies}.  Stellar-like objects with $r_{1/2}
\le 0.16$ arcsec were rejected.  Spectra were obtained using the
Low-Resolution Imaging Spectrograph (Oke \etal 1995) at the W. M. Keck
Telescope in UT April 22-24, 1996. The seeing was typically $\sim$0.8
arcsec. With a slitwidth of 1.1 arcsec, a 600 l/mm grating yielded an
instrumental resolution of $\sim$4 \AA\ FWHM at 1.26 \AA\
pix$^{-1}$. However, since our objects have $r_{1/2} \le 0.5$ arcsec,
the effective resolution is estimated to be $\sim$3.1 \AA\ (see Paper
I). Total exposure times were 3000s at each of two gratings setups
(blue and red). The total spectral range is $\sim$4000-9000 \AA , with
the exact range depending on the position of the target on the
mask. The spectroscopic reduction included the usual corrections for
bias, dark current, flatfield, and cosmic rays as well as wavelength
calibration and background sky subtraction. No flux calibration was
attempted.  The final one-dimensional spectra were produced by
coadding the central 6 pixels (1.3 arcsec) for each object. The
analysis presented in this paper refers to the sub-sample of 51
compact galaxies with measured $V - I$ colors and redshift
identifications with $z<2$. We have also excluded one nearly
stellar-like object (iw3\_0817\_0556 at z=0.960) with broad
MgII$\lambda$2795,2803 emission lines (rest-frame FWHM $\sim$23\AA )
similar to those found in low-luminosity QSOs or Seyfert galaxies (see
Appendix to Paper I).

	Figure 1 shows five representative spectra. Of the 51 compact
galaxies, 6 (or 12\%) show absorption-line spectra characteristic of
elliptical and S0 galaxies (Figure 1a).  The major features displayed
in these spectra are blended stellar absorption lines in the continuum
dominated by K-giant stars, including the 4000 \AA\ Ca H + K break,
G-band, Mgb and Na D features. Except for iw4\_1173\_1391, which shows
marginal emission in [OII]$\lambda$3727, there is no evidence for any
nebular emission within the observed wavelength range. The remaining
45 galaxies (88\%) exhibit prominent nebular oxygen and/or Balmer
emission lines, and blue continua characteristic of vigorous
star-forming systems or narrow-line active galactic nuclei.  A large
fraction have spectra that resemble those of star-forming HII regions
(Figures 1b and 1c), showing narrow lines and a wide range in
excitation as evidenced by the [OIII]$\lambda$5007/H$\beta$ flux
ratio. Particularly interesting are the spectra of seven galaxies at
$z>0.7$ showing strong [OII]$\lambda$3727 lines, and a strong
continuum bluewards of [OII]$\lambda$3727 with clear
FeII~$\lambda$2600 and MgII~$\lambda$2796 absorption features (Figure
1d).  These features are characteristic of extreme local starburst
galaxies such as NGC 2415 or NGC 5253, which are undergoing a very
recent violent episode of star formation (Kinney \etal 1996). Two
other galaxies, also at $z>0.7$, have similar blue continuum but show
unusually narrow MgII$\lambda$2796 emission (restframe FWHM $\sim$3.5
\AA ) with P-Cygni line profiles, as well as strong [OII]$\lambda$3727
and [NeIII]$\lambda$3869, and weak narrow H$\gamma$ and H$\delta$
emission lines (Figure 1e).  Although the narrow lines argue against
nuclear activity in these objects, we have not been able to find
similar spectral characteristics in the UV spectra of local
starbursts.  A more quantitative analysis of the general spectroscopic
properties of the emission-line compact galaxies is presented in
Section 4.

\section{Emission-line Measurements}

	Paper I shows that the redshift distribution of compact
galaxies parallels that of typical field galaxies in the HDF and is
roughly bimodal with peaks around $z\sim 0.5$ and $z\sim 0.9$. We thus
divide our objects into `intermediate-$z$' (26 emission-line
galaxies at $z<0.7$) and `high-$z$' (19 emission-line galaxies at
$z>0.7$) samples.  The main spectral features seen in the intermediate
sample are: [OII]$\lambda$3726,3729, H$\beta$, [OIII]$\lambda$4959,
and [OIII]$\lambda$5007. Four objects at $z<0.4$ also exhibit other
emission lines such as HeI$\lambda$5876, $H\alpha$,
[NII]$\lambda$6548,6583, and [SII]$\lambda$6717,6731.  For the
high-$z$ sample ($z>0.7$), the [OII] doublet is generally the
only major feature that can be reliably measured.  The emission-line
measurements described below refer to the strongest features most
commonly observed in our spectra, i.e., [OII]$\lambda$3726,3729,
H$\beta$, and [OIII]$\lambda$5007 (hereafter referred to as [OII],
H$\beta$ and [OIII], respectively).

\subsection{Equivalent Widths}
	Equivalent widths (EW) were measured by fitting a Gaussian
function to the emission-line profiles using the SPLOT program in
IRAF\footnote{IRAF is distributed by the National Optical Astronomy
Observatories, which is operated by the Association of Universities
for Research in Astronomy, Inc. (AURA) under cooperative agreement
with the National Science Foundation}.  The continuum levels and the
range over which the fits were performed were set interactively, with
repeat measurements made in difficult cases. The effective
instrumental resolution of 3.1\AA\ resolves the [OII] doublet in
$\sim$25\% of the spectra.  In all cases, the standard deblending
routine within SPLOT was used to fit both components. The {\sc FWHM}
for the two-Gaussian fit were forced to be equal, and the distance in
wavelength between the two centroids was fixed to the theoretical
value ($2.75\times(1+z)$\ \AA ).  All lines, whether double or single,
were generally well-fitted by Gaussian profiles. EW's were measured from
direct integration of the flux given by the Gaussian fit in the rest
frame. No correction for stellar absorption was made to the measured
EW of H$\beta$. This correction amounts typically to $\sim$2-5\AA\ for
HII galaxies and spiral galaxies (Tresse \etal 1996; Kennicutt
1992). For the [OII] doublet, we coadded the flux of each line to give
a single measure of EW that can be directly compared to that commonly
measured at lower spectral resolution.  All EW measurements were
derived independently using software designed by one of us (ACP),
which directly integrated the flux in the line. Comparing both
techniques, we estimate that the average uncertainty of our
measurements is $\sim$15\%. The histograms of $EW_{[OII]}$ in the
rest-frame for the intermediate and high-$z$ samples are shown in
Figure 2a.  These values range from 5 \AA\ to 94 \AA\ with an average
of 43 \AA\ (rms = 24\AA ). Note that there is no significant
difference in the distribution of $EW_{[OII]}$ between both redshift
samples.  Using the two-sided Kolmogorv-Smirnoff test, the probability
is $\sim$82\% for the intermediate- and high-$z$ samples to be
drawn from the same parent distribution.

\subsection{Velocity Widths}
	Velocity widths ($\sigma$) were characterized as the rms
velocity dispersion of the Gaussian fit to a given line with
rest-frame wavelength $\lambda_i$, corrected for redshift and
instrumental resolution, i.e.: 

\[ \sigma_i = \sqrt{{\sc FWHM}^2_i -
(3.1{\rm \AA})^2} \times \frac{3\times 10^5~{\rm km\ s}^{-1}}{2.35
\lambda_i (1+z)} \] 

Most of the emission-lines in our spectra do not deviate
significantly from the Gaussian fit. These profiles are consistent with
those measured in high resolution, high signal-to-noise spectra of a 
similar sample of compact, narrow emission-line galaxies (Koo \etal 1995; 
Guzm\'an \etal 1996). We estimate that the lowest $\sigma_i$ value that
can be reliably measured with our instrumental resolution is $\sim$35
km s$^{-1}$ (at a 90\% confidence level). Individual $\sigma_i$
measurements were assigned a quality parameter Q related to the
signal-to-noise ratio (SNR) per \AA\ of each line. We adopted Q=1 for
SNR $\le 20$, Q=2 for $20 <$ SNR $< 40$, and Q=3 for SNR $\ge 40$.  A
total of 48 objects have at least one $\sigma_i$ measurement with Q
$>$ 1. No significant systematic difference among $\sigma_i$
measurements from various lines of the same object was found, although
we note that the velocity widths derived from [OII] are $\sim15\% \pm
8\%$ higher than those values derived from [OIII] and H$\beta$. From
the variance among different line measurements for the same galaxy,
the typical uncertainty of an individual $\sigma_i$ measurement is
$\sim$20\%. The quality code was used to derive a final $\sigma$ value
as the weighted average of the values for all available emission lines
given by the expression:

\[ \sigma \; = \; \frac{\sum_i Q_i  \sigma_i}{\sum_i Q_i} \]

The uncertainty in the final measurements is typically $\sim$15\%. 

	Figure 2b shows the histograms of $\sigma$ measurements for
the intermediate and high-$z$ samples. The $\sigma$ values range from
about 35 km s$^{-1}$ (the measurement limit) to 150 km s$^{-1}$, with
an average value of 65 km s$^{-1}$ (rms = 29 km s$^{-1}$). In contrast
to the $EW_{[OII]}$ histogram, the $\sigma$ distribution shows
significant differences with redshift in the sense that galaxies at $z
>$ 0.7 have, on average, 40\%$\pm$10\% larger $\sigma$ than those at
$z \le$ 0.7. This effect is not due to the small systematic offset of
$\sim$8\% between [OII]-based velocity widths (used exclusively for
the high-$z$ sample) and the average velocity widths adopted for the
intermediate sample.  Using the two-sided K-S test, the probability
that both distributions are drawn from the same parent distribution is
only $\sim$1\%.

\subsection{Excitation and H$\beta$ luminosities}
	A useful indicator of the physical conditions of the ionized
gas is the [OIII]$\lambda$5007 to H$\beta$ flux ratio (so-called {\sl
excitation}).  The proximity in wavelength between H$\beta$ and [OIII]
ensures that any extinction correction in this ratio is
small. Unfortunately, for some galaxies at $z <$ 0.7, one or both
lines lie close to strong sky emission lines which prevented reliable
measurements.  In total, [OIII]$\lambda$5007/H$\beta$ could be
measured for only 24 galaxies, mostly at $z \le$ 0.7. For 28 galaxies
we have also derived H$\beta$ luminosities from the measured
rest-frame equivalent widths $EW_{H\beta}$ and absolute B magnitudes
(see Terlevich \& Melnick 1981). In principle, these luminosities need
to be corrected for internal extinction and stellar absorption. For
local star-forming galaxies, these two corrections typically amount to
$\sim$0.8 dex and $\sim$0.1 dex, respectively. Since their size is
very uncertain for our sample, we will not apply any such corrections
to the observed $H\beta$ luminosities of compact galaxies. Comparison
with local samples will be made using only un-corrected values for the
nearby galaxies.

\subsection{Star Formation Rates}
	We have estimated the SFR based on $EW_{[OII]}$.  Metallic
nebular lines like [OII] are affected by the physical conditions of
the ionized gas (e.g., excitation and metallicity), and the
transformation from SFR estimated this way to SFR from H$\alpha$
fluxes (the best SFR tracer) is not straightforward.  Previous studies
for local emission-line galaxies by Gallagher \etal (1989) and
Kennicutt (1992) give expressions for such transformations that differ
by a factor $\sim$5. Gallagher \etal studied a sample of nearby blue
irregulars, while Kennicutt used a sample of nearby galaxies covering
all disk galaxy types. The disagreement between the two calibrations
may reflect the difference in extinction and reddening between
irregulars and spirals, and the different IMF and stellar models used
by Kennicutt and Gallagher \etal (Kennicutt 1992). Other factors that
may contribute to the observed difference are the possible stronger
contribution of extended diffuse ionized gas in irregulars, or
variations in the sampling of the disk for the two galaxy types.

	In order to estimate the SFR using $EW_{[OII]}$, we
have derived our own transformation between H$\alpha$ and [OII]
fluxes using a sample of local emission-line galaxies that best
resembles the typical characteristics of our sample. The derivation  of
such transformation is described in detail in the Appendix. The final
expression to estimate the SFR as a function of the observed  $EW_{[OII]}$
for compact galaxies is:

\[ SFR (M_{\odot} \; yr^{-1}) \; \approx \; 2.5 \times 10^{-12} \times
10^{-0.4 \: (M_B-M_{B\odot})} \; EW_{[OII]} \] 

\noindent This estimate is $\sim$3 times smaller
than that  derived by Kennicutt, and $\sim$1.5 times larger than  
that obtained by Gallagher \etal Absolute blue magnitudes for our 
sample are listed in Paper I. 

	Figure 2c shows the histograms of SFR for the
intermediate- and high-$z$ samples. While compact galaxies at
$z<0.7$ have SFRs $< 3$ M$_\odot$ yr$^{-1}$, those at higher redshifts
have SFRs that span a large range from 2 to 14 M$_\odot$
yr$^{-1}$. Since the distribution in $EW_{[OII]}$ is very similar for
both samples (Figure 2a), the observed difference in SFRs mainly 
reflects the fact
that we are selecting more luminous galaxies at higher redshifts.
The median luminosities in the intermediate and high redshift samples are
M$_B$ = -19.4 and -21.0, respectively (see Paper I).
In other words, the average SFR {\sl per unit luminosity} is
similar for the intermediate- and high-$z$ samples. 

\subsection{The Data}
	A complete listing of the emission-line data is given in Table
1. Column (1) lists the galaxy identification. Columns (2) and (3)
list the apparent I$_{814}$ magnitudes and redshifts given in Paper
I. The rest-frame equivalent widths of [OII], H$\beta$, and [OIII]
in \AA\ are listed in columns (4), (5) and (6), respectively. Column
(7) lists the excitation.  Velocity widths are listed in column (8) in
km s$^{-1}$. H$\beta$ luminosities in erg s$^{-1}$, uncorrected for
extinction, are listed in column (9). Star formation rates in
M$_\odot$ yr$^{-1}$ are listed in column (10). Finally, in column (11)
we list the spectral type assigned in Section 4 below.

\section{Spectroscopic Properties of Compact Emission-Line Galaxies}

	We investigate the nature of the faint compact galaxies by
comparing their spectroscopic properties in various diagnostic
diagrams with different types of emission-line galaxies. First, we
focus the analysis on two well-known diagrams: excitation
vs. luminosity, and H$\beta$ luminosity vs. velocity width. Together,
these diagrams provide insight into the physical characteristics of
the ionized gas (e.g., metallicity and internal motions) as well as
the strength of any starburst\footnote {We define ``starburst'' as a
star-forming system with a current SFR that is at least twice the
average SFR of normal disk galaxies, which implies a typical $M/L$
ratio $\le$1 $M_\odot/L_\odot$ (cf. Kennicutt, Tamblyn \& Congdon
1994). The majority of our galaxies fit this classification (see
Paper~I).}. These plots, however, are useful only for compact galaxies
at $z<0.7$, for which H$\beta$ and [OIII] lie within the observed
wavelength range. For galaxies at higher redshift, [OII] is generally
the only major feature that can be reliably measured. Since [OII] is a
good tracer of the SFR, the $EW_{[OII]}$ vs. luminosity diagram provides
a useful tool to study the star formation characteristics of both the
intermediate- and high-$z$ galaxy samples.  We also introduce a
new diagram: SFR per unit mass vs. mass. This plot discriminates among
various types of star-forming galaxies based solely on their star
formation activity, independently of their luminosity. Finally, we propose a 
broad classification scheme of the compact galaxy sample based on this 
analysis.

\subsection{The Excitation vs. Luminosity Diagram}             
	The spectroscopic properties of narrow emission-line galaxies
are generally characterized using line-ratio diagnostic diagrams
(Baldwin, Phillips \& Terlevich 1981; Veilleux \& Osterbrock 1987).
Figure 3 shows the [OIII]/H$\beta$ vs. $M_B$ diagram for the 28 
compact galaxies with measured [OIII]/H$\beta$ ratio (squares). All
but one (filled square) belong to the intermediate
sample ($z<0.7$). Note that our magnitude limit 
prevents us from observing galaxies fainter than $M_B
\sim -17$ at $z>0.4$ (see Paper I). For comparison, we also plot 
H$\alpha$-selected emission-line galaxies from the UCM local survey
(Gallego \etal 1997) and a sample of compact narrow
emission-line galaxies (CNELGs) at $z=0.1-0.7$ studied by Koo \etal
(1995) and Guzm\'an \etal (1996). 

	In the [OIII]/H$\beta$ vs. $M_B$ diagram, local emission-line
galaxies can be grouped into two different classes: starburst galaxies
and active galaxies (e.g. Salzer \etal 1989).  The first class
consists of objects where
the gas is ionized by young O and B stars, and includes 
starburst nuclei (SBN), dwarf amorphous nuclear
starbursts (DANS) and HII galaxies. The second class contains
ionization sources harder than hot main sequence stars, such as 
Seyfert galaxies and LINERS.  
Local starburst galaxies
define a continuous sequence in Figure 3 analogous to the so-called
`HII' sequence observed in the [OIII]/H$\beta$ vs. [NII]/H$\alpha$
diagram (Veilleux \& Osterbrock 1987). This sequence is
interpreted as being a variation in the metallicity content of the
ionized gas (Dopita \& Evans 1986; Stasinska 1990). Along the HII
sequence, metallicity increases with luminosity from the HII
galaxies to the SBNs.  

	Most of the compact galaxies in the intermediate sample lie in
the moderate to high-excitation regime populated by local HII galaxies
and moderate-$z$ CNELGs (i.e., log [OIII]/H$\beta > 0.3$). Direct
comparison with Dopita \& Evans (1986) models yields an average
metallicity $Z \sim0.4 Z_\odot $ for compact galaxies in this
excitation regime. This value is consistent with that derived from the
luminosity--metallicity relation for local emission-line galaxies in
the same range of luminosities (Salzer \etal 1989). A few objects have
low [OIII]/H$\beta$ ratios consistent with more evolved star-forming
systems such as local DANS and SBNs (hereafter called `starburst disk
galaxies'). The average metallicities for these objects are $Z \sim
0.8 Z_\odot$.  From the analysis of [OIII]/H$\beta$ vs. $M_B$, we
conclude that emission-line compact objects at $z<0.7$ are vigorously
star-forming galaxies covering a broad range in metallicity.

\subsection{The H$\beta$ Luminosity vs. Velocity Width Diagram}  
	For star-forming galaxies with a dominant young population,
the flux of the Balmer lines provide a reliable estimate of the age
and the strength of the on-going burst (Dopita \& Evans 1986;
Mas-Hesse \& Kunth 1991; Leitherer \& Heckman 1995). A useful diagram
to study intrinsic differences in the evolutionary state of various
types of star-forming galaxies is $H\beta$ luminosity vs. velocity
width since the starburst properties can be compared among galaxies
with similar internal kinematics, i.e., independently of any
luminosity evolution.  Figure 4 shows the $L_{H\beta}-\sigma$ diagram
for the 28 compact galaxies with reliable $L_{H\beta}$
measurements. Only 4 of these objects have $z> 0.7$ (filled
squares). For comparison, we also show the sample of local HII
galaxies studied by Melnick, Terlevich \& Moles (1988), as well as the
sample of CNELGs with $\sigma <$ 70 km s$^{-1}$ presented in Koo \etal
(1995). Since $L_{H\beta}$ measurements for both compact galaxies and
CNELGs have not been corrected for internal extinction, we have
decreased the corrected $L_{H\beta}$ values of local HII galaxies by
$\sim$0.7 dex, which corresponds to the average value of the
extinction correction for luminous HII galaxies (Gallego \etal
1997). We also plot a sample of local infrared-selected starburst disk
galaxies studied by Lehnert \& Heckman (1996). H$\beta$ luminosities
were derived from extinction-corrected H$\alpha$ luminosities, i.e.
$L_{H\beta}= L_{H\alpha}/2.86$. These values were decreased by 0.9 dex
to account for the average extinction correction applied to starburst
disk galaxies (Gallego \etal 1997). Velocity widths were derived from
their rotational velocity measurements, assuming $\sigma = 0.426
\times 2 V_{rot}/{\rm sin}i$.
 
	In the $L_{H\beta}-\sigma$ diagram, HII galaxies with
$EW_{H\beta} >$30\AA\ follow a well-defined correlation: $L_{H\beta}
\propto \sigma^5$ (Terlevich \& Melnick 1981).  CNELGs also follow the
same trend (Koo \etal 1995), while starburst disk galaxies define a
similar relation that is offset towards lower $L_{H\beta}$ by a factor
of $\sim$30 at a given $\sigma$. The distributions of local HII
galaxies and starburst disk systems define the boundaries for the
overall observed range in $L_{H\beta}$ at any given $\sigma$ of
compact galaxies. The large observed spread in $L_{H\beta}$ reflects
variations in metallicity (Terlevich \& Melnick 1981), extinction
(since no corrections have been applied), and strength of the current
burst of star formation (since the luminosity of the Balmer lines
scales directly with the SFR). Differences in the relative
contribution of turbulent and virial motions to the velocity widths in
our sample galaxies may also affect their distribution in Figure 4,
although the effect seems to be noticeable mainly in objects with
$\sigma > 60$ km s$^{-1}$ (Melnick, Terlevich \& Moles
1988).  On average at a given $\sigma$, low $L_{H\beta}$ compact
galaxies tend to have $\sim$50\% lower excitation (i.e., higher
metallicity) and $\sim$6 times lower SFRs than their counterparts with
high $L_{H\beta}$.

Roughly, half of the compact galaxies shown in this diagram have
$H\beta$ luminosities and velocity widths similar to those of extreme
star-forming HII galaxies, in agreement with the analysis of the
line-ratio diagram discussed previously. This can be interpreted as
the result of having average metallicity, extinction, and SFR
consistent with those values typical for HII galaxies.  The remaining
have $L_{H\beta}$ and $\sigma$ values indicative of being more evolved
star-forming systems with metallicity, extinction, and SFR approaching
values characteristic of local starburst disk galaxies.

\subsection{The [OII] Equivalent Width vs. Luminosity Diagram}
	The [OII] luminosity ($L_{[OII]}$) is a good tracer of the SFR
(Gallagher \etal 1989; Kennicutt 1992). In the absence of
flux-calibrated spectra, $L_{[OII]}$ can be estimated using [OII]
equivalent widths and blue luminosities (i.e., $L_{[OII]} \sim
10^{29}\ EW_{[OII]}\ L_B$; see Appendix).  The $EW_{[OII]}-M_B$
diagram thus provides direct information on the star formation
activity. Figure 5 shows the $EW_{[OII]}-M_B$ diagram for the whole
sample of distant compact objects, as well as for a representative
sample of local emission-line galaxies (Salzer \etal 1989; Gallego
\etal 1997).  The [OII] emission lines of compact galaxies are
remarkably strong (equivalent widths of $\sim$60\AA\ in the integrated
spectrum), given that these are luminous galaxies with absolute $B$
magnitudes of about $-$20. In agreement with the results derived in
the previous sections, compact galaxies in the intermediate-$z$ sample
show [OII] equivalent widths and blue luminosities consistent with
those values characteristic of local HII and starburst disk galaxies.
In particular, compact galaxies with high excitation and high
$L_{H\beta}$ at a given $\sigma$ tend to have large $EW_{[OII]}$
similar to HII galaxies.  High-$z$ compacts also show [OII] equivalent
widths and blue luminosities that overlap the observed distribution
for local vigorously star-forming galaxies.  However, although
high-$z$ compacts exhibit a similar range in $EW_{[OII]}$ to that of
compacts at intermediate-$z$, they are $\sim$1-2 mag brighter.  This
implies that, on average at the same equivalent width, compact
galaxies at $z>0.7$ have higher [OII] luminosities by a factor
$\sim$10 than those at $z<0.7$, which in turn translates into their
having $\sim$10 times higher SFR.

	The increase in the [OII] luminosity with redshift suggests an
enhancement of the SFR in compact galaxies at higher redshifts. To
assess whether this enhancement implies a significant evolution in the
star formation activity of compact galaxies, it is necessary to take
into account the selection effects at play in our sample.  Because of
the cutoffs in $r_{1/2}$, $I_{814}$ and $\mu_{I814}$ applied, the
global galaxy properties of our sample are strongly correlated with
redshift, and also among each other.  This is clearly shown in Figure
6, where we plot the distribution of $L_{[OII]}$ (or SFR) as a
function of surface brightness for the intermediate- and high-$z$
samples. Superimposed on the data points, we plot the approximate
limits of the observable window defined by our selection effects at
$z\sim0.55$ (i.e., $1.3<R_e<4$ kpc, $M_B<-18.25$, and $SB_e<22.0$),
and at $z\sim0.85$ (i.e., $1.6<R_e<5$ kpc, $M_B<-19.25$, and
$SB_e<21.2$). These are, in fact, the same boundaries described in
Paper I using the $M_B-SB_e$ diagram. Figure 6 demonstrates that
selection effects can account for the lack of compact galaxies with
low $L_{[OII]}$ in the high-$z$ sample. However, they cannot explain
why compacts with $L_{[OII]} \ge 10^{41}$ erg s$^{-1}$ are so rare in
the intermediate-$z$ sample.  The shaded region defined by the
intersection of both observable windows represents the area of the
parameter space available to our sample galaxies in the redshift range
$0.4<z<1$, approximately. If the apparent lack of high $L_{[OII]}$
systems in the intermediate-$z$ sample were due to selection effects,
then high- and intermediate-$z$ compacts within the shaded region
would show a {\sl similar} distribution.  However, the segregation in
$L_{[OII]}$ between the two samples still remains. Note that the
volumes mapped at $0.4<z<0.7$ and $0.7<z<1$ are comparable (i.e.,
$1.6\times 10^4$ Mpc$^3$ and $2.3\times 10^4$ Mpc$^3$, respectively),
so a volume-richness effect is not present. The sparsity of high
$L_{[OII]}$ compacts in the intermediate-$z$ sample, as compared to
the numbers observed at $z>0.7$, points towards a steep evolution of
the star formation activity in compact galaxies with redshift. Given
the small number of objects involved, however, it is difficult to
demonstrate with our sample that significant evolution has actually
occurred.

Previous surveys (Glazebrook \etal 1995; Cowie \etal 1995) have
pointed to the presence of vigorously star-forming galaxies at
$z\sim$1. These galaxies have unusually high [OII] luminosities
($L_{[OII]} \sim 10^{42}$ erg s$^{-1}$) that are rarely seen in field
galaxy samples at $z<0.7$ and locally.  HST observations for nine such
galaxies show compact objects with complicated morphologies such as
chains and compact blobs (Cowie \etal 1995). Spectroscopically, they show
very strong [OII] emission lines and several absorption features
identified as MgII, MgI and FeII. These objects are similar in their
structural and spectroscopic properties to the subset of seven extreme
star-forming compact galaxies at $z>0.7$ mentioned in Section 2.  
High-$z$ compact galaxies, however, do resemble nearby
vigorously star-forming systems in terms of [OII] equivalent widths
and blue luminosities (Figure 5). This suggests that the more distant
compact galaxies are not necessarily in unique evolutionary states,
but may simply be larger versions of the same starburst systems
observed at lower redshifts.

\subsection{The specific SFR vs. Mass Diagram}
	To quantify the starburst nature of the compact galaxy sample,
we introduce a new parameter: the SFR per unit mass (or {\sl specific}
SFR), which reflects the strength of the current burst of star
formation relative to the underlying galaxy mass. Masses can be
estimated via the virial theorem using the half-light radius and
velocity width as a measure of the galaxy size and gravitational
potential. For star-forming galaxies, the virial masses will very
likely underestimate the total mass, depending on galaxy type,
inclination and aperture effects, and the presence of any older,
extended stellar population (e.g., see Guzm\'an \etal 1996).  This may
introduce a bias in the comparison of the specific SFRs derived for
various galaxy types. A detailed discussion on the reliability of
virial mass estimates for our compact galaxy sample is presented in
Paper I.  For the purposes of this paper, we have made an
independent assessment of this bias by comparing the specific SFR
derived using total and virial masses for a sample of 12 nearby blue
galaxies with $-22<M_B<-15$ that best resembles the characteristics of
our compact sample. The total mass in gas and stars for a galaxy can
be estimated from the measured HI mass and optical luminosity (i.e.,
$M_{total} \sim$ 1.34 $M_{HI} + \left(\frac{M_\star}{L_\star}\right)
L_B$; Hunter \& Gallagher 1986).  Using the values for $SFR/M_{total}$
listed in Table 2 of Hunter \& Gallagher (1986), as well as published
velocity widths (Hunter, Gallagher \& Rautenkranz 1982), [OII]
equivalent widths (Gallagher \etal 1989) and half-light radii (de
Vaucouleurs \etal 1991), we derive:

\[ \frac{SFR/M_{virial}}{SFR/M_{total}} \; = \; 1.3 \pm 0.1 \]

	Since we have used the same calibration for the SFR, this
comparison implies that virial masses are underestimating the total
masses for this sample of very blue, star-forming galaxies by
$\sim$30\% $\pm$ 15\% (i.e., less than a factor $\sim$2 at the most). This
value does not depend significantly on whether only high- or
low-luminosity galaxies are considered in the comparison. We thus
conclude that the uncertainty introduced by using virial mass
estimates to determine the $SFR/M$ of blue star-forming systems, such
as our compact galaxies, is typically less than a factor $\sim$2 over
a range of $\sim$7 mag in luminosities.

 	Figure 7 shows the specific SFR (normalized to 10$^{11}
M_\odot$) versus mass for the whole sample of compact galaxies as well
as for the sample of HII galaxies, CNELGs and starburst disk galaxies
described in Figure 4. Also for comparison, we include two other
samples of nearby blue galaxies (Gallagher \etal 1989) and luminous
starbursts (Calzetti 1997) with kinematic data.  Virial mass
estimates for our compact galaxies are listed in Paper I. For HII
galaxies, masses were derived using the half-light radius and $\sigma$
measurements listed in Telles (1995), assuming the same structural
constant adopted for our compact objects.  The SFRs of
HII galaxies were derived using $L_{H\alpha}$ (see Appendix), assuming
$L^{obs}_{H\alpha}\ E(L_{H\alpha})= 2.86 \times L_{H\beta}$, where
$E(L_{H\alpha})$ is the correction for extinction in $L_{H\alpha}$,
and $L_{H\beta}$ are the extinction-corrected luminosities given in
Melnick, Terlevich \& Moles (1988). For CNELGs, masses were taken from
Guzm\'an \etal (1996), while SFRs were derived from $EW_{[OII]}$
measurements (Koo, unpublished).  Only three such objects have
observations of [OII]. Masses and star formation rates for Lehnert \&
Heckman's sample of local starburst disk galaxies were derived from
their published values of $R_e$, $V_{rot}/{\rm sin}i$ and
$L_{H\alpha}$, assuming the gas is in dynamical equilibrium with the
same potential sensed by the disk gas, and an average extinction correction of
$\sim$0.4 dex in the observed $L_{H\alpha}$. For the blue galaxy sample, we
derive masses and SFRs from published velocity widths (Hunter,
Gallagher \& Rautenkranz 1982), [OII] equivalent widths (Gallagher
\etal 1989) and half-light radii (de Vaucouleurs \etal 1991). Finally,
the data for nearby luminous starbursts is taken directly from
Calzetti (1997).

	The $SFR/M$ vs $M$ diagram is particularly useful to analyse
the nature of compact galaxies using the star formation activity as
the only measure of the evolutionary state of galaxies with the same
mass (instead of luminosity, which depends itself on evolution). In
this diagram, HII galaxies and CNELGs describe a well-defined sequence
showing relatively constant values of the specific SFR at the top of
the overall galaxy distribution. The scatter is only $\sim$0.25 dex in
$SFR/M$ over two orders of magnitude in mass.  Most local starburst
disk systems likewise exhibit relatively constant specific SFRs, but
at a rate $\sim$30 times smaller than that of HII galaxies. These two
galaxy types define clear boundaries to the distribution of compact
objects. The distribution of compacts in the $SFR/M-M$ diagram
is indeed very similar to that observed in the $EW_{[OII]}-M_B$
diagram. This supports our assumption that any systematic bias in the
mass estimates among compact galaxies is small compared to the
intrinsic differences in their star formation activity. 

	There are two major results that can be directly drawn from
Figure 7. First, the highest values of the specific SFR exhibited by
compact galaxies are consistent with those characteristic of local HII
galaxies. Thus we do not find evidence for an increase in the peak of
the specific SFR activity with redshift in our sample. Second, compact
galaxies at $z>0.7$ are, on average, $\sim$10 times more massive
than their counterparts with similar specific SFR at $z<0.7$ (or,
alternatively, they have $\sim$10 higher $SFR/M$ at a given mass).
This effect is partly due to the magnitude cutoff adopted in our
sample selection, which translates into an offset in the low-end of
the mass distribution for both samples.  As shown in Figure 6, this
cutoff implies a difference of $\sim$1 mag between the
absolute blue magnitude limits at $z\sim0.55$ and $z\sim0.85$. Assuming
similar mass-to-light ratios for compact galaxies in both samples, the
above magnitude difference yields an offset of $\sim$0.4 dex in the
lower limit of the expected mass distributions for the high- and
intermediate-$z$ samples, which is consistent with that observed.

	Although selection effects may account for the lack of
low-mass compact objects at high redshift, they cannot explain why the
massive star-forming systems are not present in the intermediate
redshift sample.  Figure 8 shows the $SFR/M$ vs $M$ diagram for the
sub-sample of compact galaxies that could have been observed over the
whole range of redshifts $0.4<z<1$, as discussed in Section
4.3. Superimposed on the data points we have plotted the approximate
boundaries of the shaded region shown in Figure 6.  If the apparent
lack of massive star-forming systems at $z < 0.7$ were due to
selection effects, then compact galaxies in the intermediate- and
high-$z$ samples should show a similar distribution in this
figure. However, despite the small number of data points, the 
segregation in mass between the two samples is still noticeable. This
is the same effect seen in the distribution of [OII] luminosities and
described in Section 4.3. The enhancement of the SFR in compact
galaxies with redshift suggested by the analysis of $L_{[OII]}$ is not
thus the result of high-$z$ compacts having unusually high star
formation activity per unit mass. Rather, compacts at $z>0.7$ have
similar specific SFR to those at $z<0.7$, but are {\sl intrinsically
more massive}. These results agree qualitatively with the
``downsizing'' scenario discussed by Cowie \etal (1996), in which more
massive galaxies form at higher redshift. Instead of indicating a new
population of massive star-forming systems, Figure 7 shows that these
high-$z$ compact galaxies may be related to local luminous starbursts
(such as NGC 4385, NGC 4194, or NGC 1614) which, in some aspects, can
also be considered high-luminosity counterparts of less luminous HII
galaxies (cf. Stasinska \& Leitherer 1996). These results thus suggest an
increase in the {\sl number of massive star-forming systems} with
redshift.  To assess whether significant evolution has actually
occurred requires information on the relative numbers of field compact
emission-line galaxies as a function of mass in distant and nearby
galaxy samples. A preliminary attempt to tackle this issue with the
currently available data will be presented in Section 5.

	Based on these results, we have classified the compact galaxy
sample into two classes. Compacts with log$(SFR/M) < 1.25$ (i.e., the
lower limit for the observed distribution of nearby HII galaxies) are
classified as HII-like galaxies. Compacts with $SFR/M$ smaller than
this value are classified as starburst disk-like galaxies. The average
global properties of each class are summarized in Tables 2a and 2b. On
average, HII-like compact galaxies have smaller half-light radii,
larger $EW_{[OII]}$, higher surface brightnesses and SFR's per unit
mass, and lower M/L ratios than their starburst disk-like
counterparts. Within the same type, however, high-$z$ compact
galaxies have, on average, larger radii, higher luminosities, larger
velocity widths, and are more massive than those at intermediate
redshifts. This is consistent with the idea that they are simply
larger versions of the same kind of starburst galaxies found at lower
redshifts.  This broad classification of compact objects into HII-like
and starburst disk-like galaxies is also supported by slight
morphological differences seen in HST images.  HII-like objects show
generally very compact, amorphous morphologies, similar to the local
HII galaxies (Telles 1995). Starburst disk-like galaxies instead
appear to be more diffuse systems, sometimes with a central
condensation. This is the expected appearance of local starburst disk
galaxies, such as DANS or SBN, placed at intermediate redshifts.

	In summary, distant emission-line compact galaxies are similar
to local starburst galaxies. About 60\% of our sample have
luminosities, excitations, velocity dispersions, half-light radii,
surface brightnesses, mass-to-light ratios, [OII] equivalent widths,
and specific SFRs consistent with being HII galaxies, while the
remaining 40\% are more similar to starburst disk galaxies. There is no
evidence for unusually high values of the specific SFR in our
sample. However, we find that high $SFR/M$ compact objects at
$z>0.7$ are $\sim$10 times more massive than their lower redshift
counterparts. The implications for the evolution of the SFR density of
the universe will be discussed in the next section.

\section{The Evolution of the Star Formation Rate Density with Redshift}

One important goal in observational cosmology is to determine the star
formation history of the universe, and its variation with galaxy type
and environment.  Recent advances in this active field include new
observational data covering a wide range in redshift and improved
models on chemical cosmic evolution. The recent UCM survey by Gallego
\etal (1995) has determined the first reliable H$\alpha$ luminosity
function for the local universe.  The integration over all H$\alpha$
luminosities yields a total SFR density of 0.004 M$_\odot$/yr/Mpc$^3$
(using Salpeter IMF and H$_0$=50 km s$^{-1}$ Mpc$^{-1}$). This value
is almost identical to the estimate derived by Tinsley \& Danly (1980)
using colors for field spiral galaxies.  Compared to the total mass
density of stars, the SFR observed by Gallego \etal is high enough to
make all the stellar content of galaxies observed by the APM survey
(Loveday \etal 1992) in about 10 Gyr. However, the luminosity density
inferred from the APM sample could be too low by a factor of two
(Ellis \etal 1996). In that case the average past SFR would need to be
twice the currently observed value, which seems consistent with the
detection of strongly enhanced star formation in intrinsically faint
galaxies at redshifts beyond $z \sim 0.3$ (Ellis \etal 1996; Cowie
\etal 1995; Lilly \etal 1995). These results suggest a peak in the
global SFR at $z\sim$1-2 (White 1996). Such a trend has been predicted
by models of cosmic evolution using different cosmologies (Kauffmann,
Guiderdoni \& White 1994; Pei \& Fall 1995).

	Figure 9 shows a current overall picture of the evolution of
the SFR density with redshift. Published estimates were transformed to
Salpeter IMF, and $H_0$=50 km s$^{-1}$ Mpc$^{-1}$, $q_0$=0.5
cosmology, which are the values most commonly adopted in the various
studies quoted in Figure 9.  The open circles at $z=0$ are the local
values derived for the UCM sample (Gallego \etal 1995).  The filled
triangles at intermediate redshifts are SFR densities for the CFRS as
derived from 2800 \AA, 4400 \AA, and 1 \micron \ continuum
luminosities (Lilly \etal 1996).  The stars represent densities
obtained using our own $L_{[OII]}$-SFR calibration for the data set of
Cowie \etal~(1995), which is $\sim$90\% complete down to
I$_{AB}<22.5$.  Note that the value at $z\sim1.2$ may be largely
affected by incompleteness.  As an additional constraint for this
epoch, we used the upper limits obtained by Mannucci \etal (1996) in
their search for primeval galaxies at high redshift.  A lower limit
density at z$\sim$3 is given by the UV luminosities of
spectroscopically confirmed galaxies (5 initial objects from Steidel
\etal 1996 plus 11 new ones from Lowenthal \etal 1996).  The vertical
arrows at redshifts $z\sim$3.2 and $z\sim$4 are lower limit estimates
from Madau \etal (1996) using photometric redshifts for HDF. The
long-dashed horizontal line depicts the fiducial density given by the
average luminosity density observed today (Madau \etal 1996)
translated to mass and divided by the present age of the
universe. Finally, we also plot the Pei \& Fall (1995) model
predictions for three different scenarios: closed-box (C), inflow of
metal-free gas (I), and outflow of metal-enriched gas (O).

The interpretation of Figure 9 should be approached with caution,
given the likely differences in the calibrations for the various
SFR tracers, incompleteness of the data sets, and uncertainties in the
models. Despite these caveats, most of the results summarized in
this figure point to a consistent result: the total SFR density of the
universe decreased by a factor of $\sim$10 from $z=1$ to the
present-day. The nature of the galaxies responsible for this  
enhancement of the SFR density with look-back time remains an important
unanswered question.   

Although incomplete, our sample is still useful to constrain the role
of compact galaxies on the evolution of the SFR density at redshifts
$z<1$.  Instead of computing the SFR density extended to the whole
volume covered, we restrict our analysis to two well-sampled regions
in redshift space. The first one is centered at $z=0.55$, with
redshifts in the range $z =0.4$ to 0.7 (N=22 objects). The second one
is centered at $z=0.85$ and ranges from $z=0.7$ to 1.0 (N=16 objects).
For comparison with previous studies, we adopt q$_0=0.5$.  Using the
expression obtained in Section~3.2, a straightforward summation of the
observed [OII] luminosity within each redshift interval yields
co-moving SFR densities of 0.001 M$_\odot$ yr$^{-1}$ Mpc$^{-3}$ at
z=0.55, and 0.002 M$_\odot$ yr$^{-1}$ Mpc$^{-3}$ at z=0.85.  These
values correspond to 25\% and 50\% of the local estimate,
respectively.  Note, however, that only 63 of a total of $\sim$245
(i.e., 25\%) initial candidates were actually observed (see Paper
I). Assuming our sample is representative of the general population of
compact galaxies, we estimate that the total SFR densities associated
to this class are: 0.004 M$_\odot$ yr$^{-1}$ Mpc$^{-3}$ at z=0.55, and
0.008 M$_\odot$ yr$^{-1}$ Mpc$^{-3}$ at z=0.85. These values are
plotted as filled circles in Figure 9.

	A meaningful comparison of the SFR density derived for distant
compact galaxies with that of the local universe requires an estimate
of the amount of SFR for a {\sl similar} sample of nearby compact
galaxies.  The point at $z=0$ labelled as `Interm-$z$' in Figure 9
corresponds to the local SFR density derived using only 19 UCM
galaxies that satisfy the same cutoffs in $M_B$, $SB_e$, and $R_e$ 
applied to compact galaxies at
$z\sim 0.55$ (see Section 4.3).  The point at $z=0$ labelled as
`High-$z$' corresponds to the local SFR density derived for only 2 UCM
galaxies that satisfy the selection cutoffs at $z\sim$0.85.  Given
this small number, we will not consider the high-$z$ compacts in our
analysis. Direct comparison between the SFR densities derived for
similar samples of nearby and intermediate-$z$ compact galaxies
implies an enhacement of a factor $\sim$10 in the star formation
history of the universe from $z\sim0$ to $z\sim0.55$. However, this result
should be considered with caution, given the small number of galaxies
involved.

	The SFR densities derived for compact galaxies can also be
compared to the values derived for more complete galaxy samples at
similar redshifts. To avoid systematic differences due to the
calibration of the SFR using other tracers, we chose those values
derived using the $EW_{[OII]}$ data of Cowie \etal as the reference
SFR density. According to Figure 5, the global SFR density for
star-forming galaxies at $0.4<z<1$ is 0.022 M$_\odot$ yr$^{-1}$
Mpc$^{-3}$.  The Cowie \etal data refer only to field galaxies with
I$_{AB}<22.5$. For comparison, we adopt a similar magnitude limit in
our analysis. There are 301 galaxies with I$_{AB}<22.5$ in the
flanking fields of HDF. The number of compact galaxies in this sample
is 102, of which we have observed 21. Since only 14 of these compact
galaxies are emission-line systems, we estimate that the fraction of
emission-line compact galaxies is $\sim$20\% of the general field
galaxy population. Of the 14 emission-line compacts, only 12 are at
$0.4<z<1$. The SFR density for this subsample is 0.01 M$_\odot$
yr$^{-1}$ Mpc$^{-3}$. Thus, we conclude that emission-line compact
galaxies, though only $\sim$20\% of the general field population,
contribute about $\sim$45\% to the global SFR density at
$0.4<z<1$. This supports an evolutionary scenario where compact
galaxies experience major evolution at moderate redshifts. Larger, 
homogeneous data sets, that include a wide variety of structural,
kinematical, and stellar population properties for faint field galaxies,
are needed to confirm the important role that compact stellar systems 
may have in the star formation history of the universe at $z\leq1$.
\section{Conclusions}

	We have studied the spectroscopic properties of a
representative sample of 51 compact galaxies to $z<1.4$ and
I$_{814}<23.74$ in the flanking fields of the HDF. The main results
of this paper can be summarized as follows:

\begin{itemize}
	
	\item[(i)] 88\% of the spectra exhibit narrow emission-lines while
the remaining 12\% have only absorption lines. 

	\item[(ii)] Emission-line profiles are roughly Gaussian with
velocity profiles that range from $\sigma\sim 35$ km s$^{-1}$ (the
measurement limit) to 150 km s$^{-1}$. Rest-frame [OII]$\lambda$3727
equivalent widths range from 5\AA\ to 94\AA , yielding star
formation rates of $\sim$0.1 to 14 M$_\odot$ yr$^{-1}$.  

	\item[(iii)] Emission-line compact galaxies are very similar to
local starburst galaxies. About 60\% have 
velocity widths, excitations, H$\beta$ luminosities, and star formation rates
characteristic of young star-forming HII galaxies. The remaining
40\% form a heterogeneous class of more evolved starburst disk 
galaxies. 

	\item[(iv)] The distribution of specific SFRs in emission-line 
compact galaxies  at $z<0.7$ is very similar to that observed at $z>0.7$. 
The highest values of specific SFR derived for our sample are consistent 
with those characteristic of local HII galaxies.  
Thus we do not find evidence for an enhancement of the specific
SFR activity with redshift. 

	\item[(v)] On average at a given value of the specific SFR,
emission-line compact galaxies at $z>0.7$ are $\sim$10 times more
massive than their intermediate-$z$ counterparts. This result supports 
a general evolutionary
scenario in which the SFR density of the universe 
declines by a factor $\sim$10 from $z\sim1$ to the present-day.

	\item[(vi)] Emission-line compact galaxies account for only $\sim$20\%
of the general field galaxy population with I$_{AB} < 22.5$ but 
contribute $\sim$45\% to the total SFR density at
$0.4<z<1$. We conclude that compact galaxies play a significant role in the
evolution of the star formation history of the universe at moderate
redshifts. 

\end{itemize}

\acknowledgements

We thank Caryl Gronwall for providing the model SEDs used for
$k$-corrections and Katherine Wu for useful comments.  
We are grateful to the staff of the W.M. Keck Observatory
for their help during the observations. We also thank Gregory Bothun
and an anonymous referee for helpful comments which contributed to
improve the quality of this paper. Funding for this work was 
provided by NASA grants AR-06337.08-94A, AR-06337.21-94A,
GO-05994.01-94A, AR-5801.01-94A, and AR-6402.01-95A from the Space
Telescope Institute, which is operated by AURA, Inc., under NASA
contract NAS5-26555. We also acknowledge support by NSF grants AST
91-20005 and AST 95-29098. JG acknowledges the partial financial
support from Spanish MEC grants PB89-124 and PB93-456 and a UCM del
Amo foundation fellowship. JDL acknowledges funding from the Hubble
Fellowship grant HF-1048.01-93A. This research has made use of the
NASA/IPAC Extragalactic Database (NED) which is operated by the Jet
Propulsion Laboratory, Caltech, under contract with NASA.

\subsection{Appendix}

	Although the nebular H$\alpha$ luminosity is among the best
direct measurements of the current SFR (modulo the IMF), this emission
line is visible in the optical ($\lambda < 0.9\mu$) 
out to only $z\sim$0.4. At higher
redshifts, the main indicators used for estimating the SFR are
diverse: [OII] $\lambda$3727 luminosity (Dressler \& Gunn 1982;
Gallagher \etal 1984, 1989;  Broadhurst \etal 1992;
Kennicutt 1992; Cowie \etal 1995; Ellis \etal 1996), continuum colors
(Tinsley \& Danly 1980; Lilly \etal 1996) and rest-frame UV
luminosities (Steidel \etal 1996; Madau \etal 1996; Lowenthal \etal
1996). Since most emission-line compact galaxies have reliable
$EW_{[OII]}$ measurements, we chose to characterize their star formation
activity using $EW_{[OII]}$.  In order to estimate the SFR, 
we have derived our own transformation between H$\alpha$ luminosity
and $EW_{[OII]}$ using a sample of local emission-line galaxies that
best resembles the characteristics of compact galaxies. Figure 10
shows the log($EW_{[OII]}$/$EW_{[OIII]}$) vs. log([OIII]/H$\beta$) diagram for
galaxies in the intermediate compact sample compared to a
representative sample of starburst galaxies from Terlevich \etal
(1991). Most of our objects have log($EW_{[OII]}$/$EW_{[OIII]}$) 
$>$ -0.25. Thus,
as a control sample, we considered only those local galaxies that
satisfy the above condition. With this local sample of 136 objects, we
follow a similar procedure to that described by Kennicutt (1992). Assuming a
standard Salpeter IMF with m$_L$=0.1 M$_\odot$ and m$_U$=125
M$_\odot$, the SFR is related
to the observed H$\alpha$ luminosity by the expression (Alonso-Herrero
\etal 1996):

\[ SFR (M_{\odot} yr^{-1}) \; = \; 3.2 \times 10^{-42} \; L^{obs}_{H\alpha}
\; E_{H\alpha} \]

\noindent where E(H$\alpha$) is the extinction correction in the 
H$\alpha$ luminosity as derived from the Balmer decrement.  
Note that the coefficient of the new relation is $\sim$3 
times smaller than that derived by Kennicutt. This difference is 
largely due to the use of a different IMF (Alonso-Herrero \etal 1996).
Thus, in terms of the observed [OII] luminosity: 

\[ SFR (M_{\odot} yr^{-1}) \; = \; 3.2 \times 10^{-42} \; \left(
\frac{L_{H\alpha}}{L_{[OII]}} \right)_{obs} \; L^{obs}_{[OII]} \; E_{H\alpha} \]

\noindent For the local sample:

\[ \left\langle\left( \frac{L_{H\alpha}}{L_{[OII]}} \right)_{obs} \; \times \;
E_{H\alpha}\right\rangle=7.42 \pm 1.23 \]  

\noindent Adopting the average value as representative for this type
of galaxy:

\[ SFR (M_{\odot} yr^{-1}) \; = \; 2.4 \times 10^{-41} \; L_{[OII]} \]

\noindent The above coefficient is $\sim$2 times smaller than the average
calibration derived by Kennicutt for spiral galaxies.  For
very blue galaxies, however, 
such as those in our control sample and in the compact galaxy sample, 
his estimates are expected to be as much as a factor of three lower 
(Kennicutt 1992), which is consistent with our value. 
Finally, we derive the transformation between 
[OII] luminosity and $EW_{[OII]}$ using a subsample of 75 galaxies in the
control sample with known absolute blue magnitudes,
[OII] fluxes, and EW[OII] (Terlevich \etal 1991). 
The least-squares fit to the data yields:

\[ \frac{L_{[OII]}}{EW_{[OII]}} \; = \; 1.03 \pm 0.11 \times 10^{29} \; L_B(L_\odot)\]

\noindent This relation is linear over the whole range in absolute
magnitude exhibited by our control sample (i.e., $-21 < M_B < -16$). 
Substituting $L_{[OII]}$ according to this relation, the general 
expression to estimate SFR as a function of the observed EW[OII] is:

\[ SFR (M_{\odot} \; yr^{-1}) \; \approx \; 2.5 \times 10^{-12} \times
10^{-0.4 \: (M_B-M_{B\odot})} \; EW_{[OII]} \] 

Note that the above calibration may not apply for compact objects with
extreme ionization (i.e., log(EW[OII]/EW[OIII]) $<$ -0.5). There are two
such objects in our sample: sw3\_1455\_0476 is a nearby low-luminosity
blue compact dwarf with a very high excitation
(EW[OII]/EW[OIII]=0.26, [OIII]/H$\beta$=5.27). The emission lines
reveal a low-metallicity object where the electronic temperature is
very high and most of the oxygen atoms are O$^{++}$. In
such situations, the SFR derived from the [OII] luminosity may be
seriously underestimated. Since H$\alpha$ is still
visible for this object, we were able to obtain a direct measurement
of the SFR from H$\alpha$. Using [OII] we
derived a SFR=0.2 M$_{\sun}$/yr while H$\alpha$ yields
1.2 M$_{\sun}$/yr. We adopt the latter value as correct. The same
situation applies to ie3\_1030\_0507 (EW[OII]/EW[OIII]=0.215,
[OIII]/H$\beta$=5.42). In this case, the EW[OII]-based SFR quoted in Table 1
may be underestimated by as much as a factor of five.

\newpage
{\bf Figure Captions}

\figcaption[f1_panel.ps]{Spectra of five objects showing a wide range 
of spectroscopic properties: (a) absorption spectrum characteristic 
of early-type galaxies; (b) emission-line spectrum characteristic of 
low-excitation HII regions; (c) emission-line spectrum characteristic
of high-excitation HII regions; (d) emission-line spectrum with strong
blue continuum and FeII and MgII absorption features similar to that
of starbursts undergoing a violent episode of star formation; (e)
emission-line spectrum with narrow MgII in emission. 
\label{f1-panel}}

\figcaption[f2_histo2.ps]{Histograms of emission-line measurements for
the intermediate- and high-$z$ samples of compact galaxies: (a) 
[OII] equivalent widths; (b) velocity widths. A significant fraction
of objects have $\sigma$ upper-limits of 35 km~s$^{-1}$ (hatched region), 
which corresponds to the measurement limit; (c) star formation rates. 
\label{f2-histo2}}

\figcaption[f6_o3mb.ps]{Rest-frame absolute blue magnitude versus 
excitation for the
intermediate sample (open squares). Only one object from the
high-$z$ sample has measurable [OIII]/H$\beta$ ratio (filled
square).  The local sample is represented by the H$\alpha$-selected
emission-line galaxies analyzed by Gallego \etal (1997), and includes
Seyfert-2 galaxies (Sy 2), HII galaxies (HII), starburst nuclei (SBN),
and dwarf amorphous nuclei starbursts (DANS).  Symbols are as
explained in the legend. A sample of compact, narrow emission-line
galaxies (CNELGs) at $z=0.1-0.7$ (Koo \etal 1995; Guzm\'an \etal 1996) 
is also shown (stars). The location of more quiescent, early-type spirals is
represented by the box.  
\label{f6-o3mb}}

\figcaption[f7_hbs.ps]{Velocity width $\sigma$ in km s$^{-1}$ versus 
observed H$\beta$ luminosity in erg s$^{-1}$. 
Symbols for the compact objects are as
in Fig~3. Infrared-selected starbursts disk galaxies (Lehnert \& Heckman 
1996) are plotted as `$\oplus$'. The arrows represent upper-limit
$\sigma$ measurements. The approximate location of spiral galaxies 
is represented by the box. The error bar is an estimate of  
the uncertainty associated to the extinction corrections characteristic 
of the various galaxy types plotted in this figure.
\label{f7-hbs}}

\figcaption[f8_sfr.ps]{Rest-frame absolute blue magnitude versus [OII] 
equivalent width 
for our sample of compact galaxies as well as a representative sample of
nearby emission-line galaxies. Symbols are as in Figure~3.
\label{f8-sfr}}

\figcaption[f8_sfr.ps]{Rest-frame average blue surface brightness in
mag arcsec$^{-2}$ versus [OII] luminosity in erg s$^{-1}$ for our
sample of compact galaxies. Symbols are as in Figure~3. 
Dashed lines represent the observable
window in parameter space defined by the selection effects at
$z\sim0.55$ as described in section 4.3. Dotted lines represent the same
observable window at $z\sim0.85$. The shaded area corresponds to the
intersection of both windows and represents approximately the region of the
parameter space that is available for observation over the redshift 
range $0.4<z<1$, as defined by the selection effects. 
\label{f8-sfr}}

\figcaption[f8_sfr.ps]{Mass in M$_\odot$ versus star formation rate per 
unit mass in yr$^{-1}$ (normalized to $10^{11}$M$_\odot$). Symbols are 
as in Figure~4. Samples of nearby Irr galaxies (Gallagher \etal 1989) and
luminous starbursts (Calzetti 1997) are also plotted. The arrows
represent upper-limits in M and lower-limits in $SFR/M$ due to our measurement
limit in $\sigma$. The dotted line shows the division between HII-like
and starburt disk-like galaxy types as discussed in section 4.4.
\label{f8-sfr}}

\figcaption[f8_sfr.ps]{Mass in M$_\odot$  versus star formation rate per 
unit mass in yr$^{-1}$ (normalized to $10^{11}$M$_\odot$) for 
a sub-sample of compact galaxies within the shaded 
area shown in Figure 6. This sub-sample corresponds to those galaxies 
that could have been observed over the entire redshift range $0.4<z<1$, 
as discussed in section 4.3. Solid lines represent  approximately the 
limits defined by our selection effects. Note the segregation in mass 
between the intermediate- and high-$z$ samples. 
\label{f8-sfr}}

\figcaption[f9_sfrz.ps]{Current SFR density as a function of redshift 
(assuming Salpeter IMF and H$_0$=50 km s$^{-1}$ Mpc$^{-1}$, 
q$_0$=0.5 cosmology). The horizontal dashed
line represents the value needed to produce the observed stars today.  
The dotted lines represent Pei \& Fall's (1995) predictions: C =
Closed-box model, I = Inflow of metal-free gas model, and O = Outflow of
metal-enriched gas model. See text in Section 5 for description of the 
various data sets.
\label{f9-sfrz}}

\figcaption[f4_exci.ps]{Excitation versus [OII]/[OIII] ratio. Open
squares are compact objects in the intermediate-$z$ sample. Filled 
squares are compact galaxies in the high-$z$ 
sample. Crosses show the sample of starburst systems from  
Terlevich {\it et al~'}s (1991) atlas of HII galaxies. 
\label{f4-exci}}

\end{document}


\pagestyle{empty}
\small
\tablecaption{Spectroscopic parameters for HDF compact galaxies.}
\begin{deluxetable}{ccccccccccl}
\tablewidth{0pc}
\tablenum{1a}
\tablehead{
Object & I$_{\rm 814}$ & z & W$_{[OII]}$ & W$_{H\beta}$ &   
 W$_{[OIII]}$ & \raisebox{-0.1cm}{$\frac{[OIII]}{H\beta}$} & $\sigma$ & 
L$_{H\beta}$ &
 SFR & Type 
\nl
\colhead{(1)} & \colhead{(2)} &  \colhead{(3)} & \colhead{(4)} &
\colhead{(5)} & \colhead{(6)} & \colhead{(7)} & \colhead{(8)} & 
\colhead{(9)} & \colhead{(10)}& \colhead{(11)}
}
\startdata
  sw3\_1455\_0476 & 21.79 & 0.137 &    66  &    48 & 250    &  5.27 & $<$35 & 
40.24 &  1.2 & HII-like\nl
  se3\_0982\_0226 & 23.44 & 0.224 &    20  &\ldots & 11     & \ldots& 41   
&\ldots& 0.02& SB disk-like\nl
  ow3\_1047\_0444 & 22.91 & 0.318 &    37  &     9 &   30   &  2.93 & 37   & 
39.84 &  0.2 & HII-like\nl
  se3\_1090\_0023 & 21.83 & 0.410 & \ldots &\ldots & \ldots & \ldots&\ldots 
&\ldots &\ldots& E/S0 \nl
  ne3\_1126\_0494 & 22.66 & 0.420 &    30 &      8 & 22     & 3.33   & 51& 40.12 
& 0.3 & SB disk-like\nl
  oe2\_0040\_0229 & 22.78 & 0.423 & 60  &    15 & 25     &  1.91 & 78    & 40.31 
&  0.5 &SB disk-like \nl
  iw4\_1499\_0841 & 23.02 & 0.437 &    43 &     10 & 29     & 3.02  &  37 & 
40.12 &   0.4 & HII-like\nl
  nw4\_1111\_0965 & 22.44 & 0.457 &    46 &     10 & 12     & 1.22   & 63    & 
40.41 & 0.8 & SB disk-like \nl
  iw2\_0106\_0094 & 23.09 & 0.462 &    79 &     21 & 123    & 2.99   & 45    & 
40.47 &   0.7 & HII-like\nl
  ie3\_1071\_0351 & 22.98 & 0.476 &    35 &     10 &  9     & 1.79   & $<$35 & 
40.22 &   0.4 & HII-like\nl
  oe2\_0513\_0235 & 21.09 & 0.480 &    24  &     7 &  4     &  0.59 & 97    & 
40.76 &  1.3 &  SB disk-like \nl
  sw4\_1068\_1043 & 21.90 & 0.482 &    42  &     9 &  24    &  2.75 & 45    & 
40.64 &  1.3 & HII-like\nl
  oe4\_1100\_1320 & 23.02 & 0.483 &    53  &     9 &   33   &  4.35 & 43    & 
40.20 &  0.6 & HII-like\nl
  iw3\_1144\_0405 & 23.35 & 0.485 &    90 &     18 & 34     & 2.38   & 41    & 
40.32 &   0.7 & HII-like\nl
  oe3\_0801\_0291 & 21.97 & 0.487 &    38  &     9 &   17   &  2.12 & 69    & 
40.59 &  1.0 &  SB disk-like \nl
  nw2\_0557\_0629 & 22.15 & 0.503 &\ldots & \ldots & \ldots & \ldots &\ldots 
&\ldots &\ldots & E/S0 \nl
  iw2\_0186\_0458 & 22.41 & 0.507 &    41 &     10 & 30     & 2.94   & $<$35 & 
40.51 &   0.9 & HII-like\nl
  sw3\_1378\_0453 & 21.12 & 0.509 &     5  &     5 & \ldots & \ldots& $<$35 & 
40.67 & 0.3 & SB disk-like \nl
  se4\_0940\_1478 & 22.62 & 0.513 &    60  &    26 & 46     & 2.16  & 47    & 
40.85 &  1.1 & HII-like\nl
  iw4\_1173\_1391 & 20.84 & 0.519 &\ldots & \ldots & \ldots & \ldots &\ldots 
&\ldots &\ldots & E/S0 \nl
  ow3\_1063\_0391 & 23.23 & 0.528 &    35  &    18 &   29   &  1.43 & 50    & 
40.47 &  0.4 & SB disk-like \nl
  nw4\_1278\_1205 & 20.57 & 0.558 &\ldots & \ldots & \ldots & \ldots &\ldots 
&\ldots &\ldots & E/S0 \nl
  ie4\_1442\_1007 & 21.13 & 0.560 &    11 &      7 & \ldots & \ldots & 117   & 
40.89 &   0.8 & SB disk-like \nl
  ne3\_0969\_0527 & 22.19 & 0.570 &     8 & \ldots & \ldots & \ldots & 79    
&\ldots &   0.2 & SB disk-like \nl
  nw4\_0920\_1247 & 22.21 & 0.570 &    48 &      9 & 24     & 5.52 & $<$35 & 
40.65 &   1.6 & HII-like\nl
  ie3\_0998\_0401 & 21.81 & 0.594 &    41 &     18 & 19     & 1.13   & 91    & 
41.16 &   2.1 & SB disk-like \nl
  se2\_0673\_0133 & 22.44 & 0.594 &    48  &    18 &  53    &  2.64 & 68    & 
40.92 &  1.4 & SB disk-like \nl
  ow3\_1308\_0563 & 23.73 & 0.654 &    87  &    18 &   67   &  3.37 & 63    & 
40.47 &  1.0 & HII-like\nl
  ne3\_1323\_0484 & 21.69 & 0.677 &\ldots & \ldots & \ldots & \ldots &\ldots 
&\ldots &\ldots & E/S0 \nl
  ow3\_1376\_0297 & 22.28 & 0.693 &    35  &    10 &   14   & 0.80: & 44 & 40.86 
&  1.7 & HII-like\nl
  ie3\_1030\_0507 & 23.44 & 0.693 &    69 &    132 & 321    & 5.42   & 47    & 
41.53 &   1.2 & HII-like\nl
\enddata
\tablenotetext{}{
N{\sc OTE}.--- 
(1) Object ID: (field,CCD)\_(x)\_(y), with x and y in pixels; 
(2) Total I$_{814}$ magnitude zero-pointed to Vega; errors are
estimated to be $\pm$0.05mag (see Paper I); 
(3) Redshift, with uncertainty $\lesssim$0.001; 
(4) Rest-frame [OII]$\lambda$3727 equivalent width in \AA \ (errors $\sim$15\%);  
(5) Rest-frame H$\beta$ equivalent width in \AA \ (errors $\sim$15\%); 
(6) Rest-frame [OIII]$\lambda$5007 equivalent width in \AA \ (errors 
$\sim$15\%); 
(7) Observed excitation parameter [OIII]/H$\beta$ (errors $\sim$25\%); 
(8) Line-width (gaussian sigma), in restframe km/s (typical
uncertainty $\sim$20\%); 
(9) Log of the observed H$\beta$ luminosity in erg/s ($\pm$0.1 uncertainty); 
(10) Total SFR in M$_{\sun}$ yr$^{-1}$ (errors $\sim$15\%); 
(11) Spectroscopic type, with HII-like being those with 
log$(SFR/M)$ $>$ 1.25 and SB disk-like those with log$(SFR/M)$ $<$ 1.25. 
The values marked with ':' are very uncertain ($\sim$50\%).
}
\label{table1a}
\end{deluxetable}

\tablecaption{Spectroscopic parameters for HDF compact galaxies. Continued.}
\begin{deluxetable}{ccccccccccl}
\tablewidth{0pc}
\tablenum{1b}
\tablehead{
Object & I$_{\rm 814}$ & z & W$_{[OII]}$ & W$_{H\beta}$ &   
 W$_{[OIII]}$ & \raisebox{-0.1cm}{$\frac{[OIII]}{H\beta}$} & $\sigma$ & 
L$_{H\beta}$ &
 SFR & Type 
\nl
\colhead{(1)} & \colhead{(2)} &  \colhead{(3)} & \colhead{(4)} &
\colhead{(5)} & \colhead{(6)} & \colhead{(7)} & \colhead{(8)} & 
\colhead{(9)} & \colhead{(10)}& \colhead{(11)}
}
\startdata
  nw3\_1434\_0108 &22.82 & 0.744 &    33 & \ldots & \ldots & \ldots & 112 
&\ldots & 1.1 & SB disk-like \nl
  sw4\_0807\_1067 &22.24 & 0.760 &    34  &    10 &  21    &  2.58 & 61    & 
40.99 &  2.1 & HII-like\nl
  oe3\_1089\_0455 &21.84 & 0.784 &    33  &     9 & \ldots & \ldots& 64    & 
41.13 &  3.3 &  HII-like\nl
  ie3\_1025\_0359 &23.49 & 0.788 &    80 &     20 & \ldots & \ldots & 53    & 
40.83 &   1.8 & HII-like\nl
  oe4\_1223\_1134 &22.05 & 0.821 &    93  &\ldots & \ldots & \ldots& 68    
&\ldots &  8.6 &  HII-like\nl
  ne3\_0769\_0583 &21.60 & 0.840 &    74 & \ldots & \ldots & \ldots & 109   
&\ldots &  11.0 & HII-like\nl
 se2\_0021\_0728a &22.20 & 0.845 &    27  &\ldots & \ldots & \ldots& 65    
&\ldots &  2.3 & SB disk-like \nl
  nw4\_0770\_1239 &22.13 & 0.847 &    45 & \ldots & \ldots & \ldots & 72    
&\ldots &   4.3 & HII-like\nl
  nw2\_0367\_0766 &21.00 & 0.873 &\ldots & \ldots & \ldots & \ldots &\ldots 
&\ldots &\ldots & E/S0 \nl
  iw4\_1212\_1333 &23.55 & 0.880 &    81 & \ldots & \ldots & \ldots & 53    
&\ldots &   2.3 & HII-like\nl
  ne4\_1136\_1005 &22.94 & 0.898 &    32 & \ldots & \ldots & \ldots & 65    
&\ldots &   1.8 & HII-like\nl
  se4\_1387\_0990 &22.78 & 0.905 &    43  &\ldots & \ldots & \ldots& 109   
&\ldots &  2.9 & SB disk-like \nl
  ie3\_1443\_0559 &21.91 & 0.907 &    94 &     29 & \ldots & \ldots & 61    & 
41.81 &  13.5 & HII-like\nl
  oe2\_0168\_0249 &22.23 & 0.909 &    15  &\ldots & \ldots & \ldots& 69    
&\ldots &  1.8 &  SB disk-like \nl
  oe4\_1109\_1148 &21.47 & 0.936 &    54 & \ldots & \ldots & \ldots & 151   
&\ldots &  13.0 & SB disk-like \nl
  se2\_0309\_0594 &21.79 & 0.936 &    10  &\ldots & \ldots & \ldots& 128   
&\ldots &  2.1 & SB disk-like \nl
  ne4\_1292\_0943 &23.29 & 1.050 &    80 & \ldots & \ldots & \ldots & 43    
&\ldots &   5.3 & HII-like\nl
  oe2\_0342\_0725 &22.74 & 1.084 &    47  &\ldots & \ldots & \ldots& 84    
&\ldots &  7.4 &  HII-like\nl
  iw4\_1002\_1353 &22.17 & 1.221 &    36 & \ldots & \ldots & \ldots & 77    
&\ldots &  14.1 & HII-like\nl
  iw2\_0497\_0674 &23.00 & 1.344 &    43 & \ldots & \ldots & \ldots & 75    
&\ldots &   8.0 & HII-like\nl
\enddata
\tablenotetext{}{
N{\sc OTE}.--- 
(1) Object ID: (field,CCD)\_(x)\_(y), with x and y in pixels; 
(2) Total I$_{814}$ magnitude zero-pointed to Vega; errors are
estimated to be $\pm$0.05mag (see Paper I); 
(3) Redshift, with uncertainty $\lesssim$0.001; 
(4) Rest-frame [OII]$\lambda$3727 equivalent width in \AA \ (errors $\sim$15\%);  
(5) Rest-frame H$\beta$ equivalent width in \AA \ (errors $\sim$15\%); 
(6) Rest-frame [OIII]$\lambda$5007 equivalent width in \AA \ (errors 
$\sim$15\%); 
(7) Observed excitation parameter [OIII]/H$\beta$ (errors $\sim$25\%); 
(8) Line-width (Gaussian sigma) in km s$^{-1}$ (typical
uncertainty $\sim$20\%); 
(9) Log of the observed H$\beta$ luminosity in erg s$^{-1}$ 
($\pm$0.1 dex uncertainty); 
(10) Total SFR in M$_{\sun}$ yr$^{-1}$ (errors $\sim$15\%); 
(11) Spectroscopic type. 
}
\label{table1b}
\end{deluxetable}